\documentclass[aip,jcp,reprint]{revtex4-1}

\usepackage{graphicx}
\usepackage{amssymb,amsfonts,amsmath}
\usepackage{epsf}
\usepackage{subfigure}
\usepackage{epstopdf}
\usepackage{dcolumn}
\usepackage{bm}
\usepackage{mathrsfs}
\usepackage{mhchem}
\usepackage{color}
\usepackage{ulem}

\newcommand{\be}{\begin{equation}}
\newcommand{\ee}{\end{equation}}
\newcommand{\bea}{\begin{eqnarray}}
\newcommand{\eea}{\end{eqnarray}}

\begin{document}

\title{\bf Multistate reversible copolymerization of nonMarkovian chains \\ under low conversion conditions}


\author{Pierre Gaspard}
\affiliation{Center for Nonlinear Phenomena and Complex Systems,\\
Universit\'e Libre de Bruxelles (U.L.B.), Code Postal 231, Campus Plaine,
B-1050 Brussels, Belgium}

\begin{abstract}
The reversible kinetics of copolymerization is solved analytically for the multistate mechanism proposed by B. D. Coleman and T. G. Fox [J. Chem. Phys. {\bf 38}, 1065 (1963)] under low conversion conditions where the concentrations of monomeric species are chemostatted and stay constant in time.  Although the rates of this mechanism only depends on the currently attached or detached monomer, the growing macromolecular chain forms a nonMarkovian sequence that is characterized by matrices associated with every monomeric unit composing the sequence.  These matrices are obtained by solving the kinetic equations and they determine the growth velocity of the copolymers, the statistical properties of its possible sequences, as well as the thermodynamics of the copolymerization process.
\end{abstract}

\vskip 0.5 cm

\maketitle

\section{Introduction}

In 1963, Coleman and Fox proposed a kinetic mechanism for the growth of polymers with nonMarkovian diastereosequence distributions.\cite{CF63JCP,CF63JACS}  Their work concerns the tacticity of polymers, i.e., the property that consecutive identical monomeric units composing a polymer may have isotactic or syndiotactic placements, possibly forming atactic sequences.\cite{CF63JPS}  In the proposed mechanism, the growing polymer has two possible reactive states, each with its own stereospecificity for the attachment of monomers.  Moreover, the attachment rates are assumed to be independent of previously incorporated monomers.\cite{CF63JCP}  If the polymer always remained in a single reactive state during the growth, the sequence would form a Bernoulli chain.  Therefore, the possibility for the reactive state of the growing polymer to randomly change between two or more reactive states modifies the statistical properties of the chain and allows correlations to appear between consecutive monomeric units, generating instead a nonMarkovian chain.  In the fully irreversible growth regime where the detachment rates are negligible in front of the attachment rates, Coleman and Fox showed that the nonMarkovian chains generated by this multistate mechanism have sequence probability distributions given by products of matrices.\cite{CF63JCP}

Now, the multistate mechanism may also be considered for coplymerization processes, in which different species of monomers are incorporated in the growing chain.  Moreover, in order to investigate the thermodynamic properties of such processes, the reversed reactions of monomeric detachments should no longer be neglected, as previously studied by numerical simulations in Ref.~\onlinecite{AG09}.
In this context, the question arises whether the analytical methods developed by Coleman and Fox\cite{CF63JCP} can be extended to reversible multistate processes.

The purpose of the present paper is to show that such an extension can indeed be achieved and that the sequence probability distributions are also given by matrix products in the general case of reversible multistate copolymerization processes under low conversion conditions.\cite{CF63JCP}  These latter are required in order to achieve the stationarity of monomer concentrations during the copolymer growth.  If stationarity did not hold, the copolymer sequence would depend on the peculiar time variation undergone by the monomer concentrations in the solution,\cite{KZMI18} a situation encountered for instance in closed reactors where the population of monomers is depleted as the copolymers are growing until full equilibrium is reached.\cite{SP85,S87,S92,BL17}  Instead, it is here supposed with Coleman and Fox\cite{CF63JCP} that the monomer concentrations stay constant in time, which is the case under low conversion conditions or in an open reactor where the solution is continuously fed with monomers.  The process is described using a stochastic approach at the level of a single copolymer.  In this approach, the time evolution is considered for the probability to find the copolymer with given sequence, length, and reactive state.  The time dependence of this probability is ruled by kinetic equations in terms of the transition rates associated with the different reactions in the process.

The content of the paper is the following.  

In Section~\ref{Kinetics}, the kinetic equations of multistate reversible copolymerization are introduced and solved in the long-time limit in terms of matrices of size equal to the number of reactive states in the mechanism.  A matrix is associated with every species of monomeric units in the growing copolymer sequence, as well as with the stationary probability distribution of the different reactive states of the copolymer.  The probability of any given sequence is thus obtained as a corresponding matrix product.  The mean growth velocity is shown to be given in terms of a velocity matrix, satisfying a self-consistent equation, which plays a central role in the theory.  Furthermore, these different matrices determine the probability of any given monomeric subsequence in the bulk of the whole sequence, as well as the mean value, the variance, and the correlation function of any observable quantity defined along the sequence.  The results previously obtained in the fully irreversible regime by Coleman and Fox\cite{CF63JCP} are recovered.  

In Section~\ref{Thermo}, thermodynamics is developed for multistate reversible copolymerization in steady growth regimes.\cite{AG08} In particular, the expression of the entropy production rate is deduced and the consequences of the equilibrium detailed balance conditions are studied.  

In Section~\ref{Examples}, theory is compared with simulations using Gillespie's algorithm\cite{G76,G77} for two illustrative examples: on the one hand, an example where the thermodynamic equilibrium limit exists and, on the other hand, one without equilibrium.  

Conclusion and perspectives are given in Section~\ref{Conclusion}.

\section{Kinetics}
\label{Kinetics}

\subsection{The kinetic equations}
\label{Equations}

We consider the Coleman-Fox multistate mechanism\cite{CF63JCP} including the reversed reactions for the copolymerization of chains composed of different species of monomeric units $m=1,2,...,M$:
\bea
m_1m_2\cdots m_{l-1}^{*} \  + \  m_l \ \ &\rightleftharpoons&\ \ m_1m_2\cdots m_{l-1}m_l^{*} , \label{R1}\\
m_1m_2\cdots m_{l-1}^{**} \  + \  m_l \ \ &\rightleftharpoons&\ \ m_1m_2\cdots m_{l-1}m_l^{**} , \label{R2}\\
m_1m_2\cdots m_{l-1}m_l^{*} \ \ &\rightleftharpoons&\ \ m_1m_2\cdots m_{l-1}m_l^{**} , 
\label{R12}
\eea
in the case of two reactive states, which are here below denoted by $i=1,2$.  The transition rates of the reactions~(\ref{R1}) and~(\ref{R2}) are $w_{i,\pm m_l}$ and those of the transitions between the two states are $w_{1\to 2}$ and $w_{2\to 1}$.  The attachment rates are proportional to the concentration $c_m$ of the corresponding monomer in the solution, $w_{i,+m}=k_{i,+m}c_m$, with some rate constant $k_{i,+m}$.  Instead, the detachment rates, as well as the rates of transitions between the two states, do not depend on the monomer concentrations, $w_{i,-m}=k_{i,-m}$ and $w_{i\to j}=k_{i j}$.  Under low conversion conditions, we may assume that the monomer concentrations $c_m$ and thus the rates stay constant in time.\cite{CF63JCP}  

The copolymerization of a single chain is described as a stochastic process in terms of the probabilities $P_t(m_1m_2\cdots m_l,i)$ to find the polymer with the different possible sequences $m_1m_2\cdots m_l$ of length $l$ and states $i$ at the time $t$.  These probabilities are ruled by the following infinite hierarchy of coupled kinetic equations,
\bea
&&\frac{d}{dt}\, P_t(m_1 \cdots m_l,i) = w_{i,+m_l} \, P_t(m_1\cdots m_{l-1},i) \nonumber\\
&&\qquad\qquad\qquad +\sum_{m_{l+1}=1}^M w_{i,-m_{l+1}} \, P_t(m_1\cdots m_{l}m_{l+1},i) \nonumber\\
&&\qquad\qquad\qquad +\ w_{j\to i} \, P_t(m_1\cdots m_{l},j)\nonumber\\
&& - \bigg( w_{i,-m_{l}} + \sum_{m_{l+1}=1}^M w_{i,+m_{l+1}}+w_{i\to j}\bigg)  P_t(m_1\cdots m_{l},i) , \nonumber\\
&& \label{eq-i}
\eea
where $j=2$ if $i=1$ or $j=1$ if $i=2$.  The total probability is conserved in time, $\sum_{\omega} P_t(\omega) = 1$, where the sum extends over the sequences $\omega=m_1m_2\cdots m_l$ and the states $i$.  

\subsection{Solving the kinetic equations}
\label{Solve}

The kinetic equations are solved by introducing the following set of probabilities:
\bea
p_t(i,l) &\equiv& \sum_{m_1\cdots m_l} P_t(m_1\cdots m_{l-2}m_{l-1}m_l,i) ,\label{p0} \\
p_t(m_l,i,l) &\equiv& \sum_{m_1\cdots m_{l-1}} P_t(m_1\cdots m_{l-2}m_{l-1}m_l,i) ,\label{p1} \\
p_t(m_{l-1}m_l,i,l) &\equiv& \sum_{m_1\cdots m_{l-2}} P_t(m_1\cdots m_{l-2}m_{l-1}m_l,i) ,\label{p2} \quad \\
&\vdots&\nonumber
\eea
Introducing the notation
\be
a_i \equiv \sum_m w_{i,+m} \, , 
\ee
these probabilities obey the following equations as a consequence of the kinetic equations~(\ref{eq-i}):
\bea
&&\frac{d}{dt}\, p_t(i,l) = a_i \, p_t(i,l-1) \nonumber\\
&&\ +\sum_{m_{l+1}} w_{i,-m_{l+1}} \, p_t(m_{l+1},i,l+1) + w_{j\to i} \, p_t(j,l)\nonumber\\
&&\  - \sum_{m_l} w_{i,-m_{l}}\, p_t(m_l,i,l) -(a_i +w_{i\to j}) \, p_t(i,l) \, , 
\label{eq-p0-i}\\
&&\frac{d}{dt}\, p_t(m_l,i,l) = w_{i,+m_l} \, p_t(i,l-1) \nonumber\\
&&\ +\sum_{m_{l+1}} w_{i,-m_{l+1}} \, p_t(m_lm_{l+1},i,l+1) + w_{j\to i} \, p_t(m_l,j,l)\nonumber\\
&&\  - (w_{i,-m_{l}}+a_i +w_{i\to j}) \, p_t(m_l,i,l) \, , 
\label{eq-p1-i}\\
&&\qquad\qquad\qquad\qquad\qquad\vdots \nonumber
\eea
and equations similar to Eq.~(\ref{eq-p1-i}) for the probabilities~(\ref{p2}) and the next ones.  We note that we recover Eq.~(\ref{eq-p0-i}) by summing Eq.~(\ref{eq-p1-i}) over $m_l$, and similarly for the further equations in the hierarchy.  

Since these equations are linear, their general solution can be written as a linear superposition of solutions of the form
\bea
&& p_t(m_{l-r+1}\cdots m_{l-1}m_l,i,l) \nonumber\\
&& = \exp(s_q t + \i ql)\,  \chi_q(m_{l-r+1}\cdots m_{l-1}m_l,i)
\label{sol}
\eea
with an arbitrary parameter $-\pi < q \leq +\pi$.  The exponential rate $s_q$ is here expected to have the following form,
\be
s_q = - \i \, q \, v -{\cal D} q^2 + O(q^3) \, ,
\label{dispersion1}
\ee
where $\i=\sqrt{-1}$, $v$ is the mean growth velocity of the copolymer chain counted in monomers per second, and $\cal D$ is the diffusivity of the random drift of the length $l$.  The particular solutions~(\ref{sol}) are substituted in Eqs.~(\ref{eq-p0-i}), (\ref{eq-p1-i}),... of the hierarchy. We carry out an expansion in powers of $q$ around $q=0$ and use the fact that $s_0=0$ together with the notation
\be
\psi(m_{l-r+1}\cdots m_{l-1}m_l,i)= \chi_0(m_{l-r+1}\cdots m_{l-1}m_l,i) \, .
\label{psi}
\ee
At order $q^0$, the following equations are obtained
\bea
&& 0 = -w_{i\to j} \, \psi(i) + w_{j\to i} \, \psi(j) , \label{eq-psi0-i}\\
&& 0 = w_{i,+m_l} \, \psi(i) +\sum_{m_{l+1}} w_{i,-m_{l+1}} \, \psi(m_lm_{l+1},i)\nonumber\\
&&\ \ - (w_{i,-m_{l}}+a_i +w_{i\to j}) \, \psi(m_l,i)+ w_{j\to i} \, \psi(m_l,j) , \label{eq-psi1-i} \quad \\
&&\qquad\qquad\qquad\qquad\qquad\vdots \nonumber
\eea
and equations similar to Eq.~(\ref{eq-psi1-i}) for the quantities~(\ref{psi}) with the next values of $r$.

At order $q^1$, Eq.~(\ref{eq-p0-i}) give
\bea
 -\i v\psi(i) &=& -\i a_i \psi(i) + \i \sum_{m} w_{i,-m} \, \psi(m,i)\nonumber\\
&& -w_{i\to j} \, \psi'(i) + w_{j\to i} \, \psi'(j)  \, , \label{eq-psi0-i-q}
\eea
where $j=2$ if $i=1$ or $j=1$ if $i=2$, $m=m_{l+1}$, and $\psi'(i)=d\chi_q(i)/dq\vert_{q=0}$.  Moreover, we suppose that the following normalization condition is satisfied:
\be
\psi(1)+\psi(2)=1 \, .
\label{norm}
\ee
Summing Eq.~(\ref{eq-psi0-i-q}) over $i=1,2$, we obtain the relation giving the mean growth velocity
\bea
v &=& \sum_{i}a_i  \, \psi(i) - \sum_{m,i} \psi(m,i) \, w_{i,-m} \nonumber\\
&=& \sum_{m,i} \left[\psi(i) \, w_{i,+m} - \psi(m,i) \, w_{i,-m}\right]  .
\label{v}
\eea

In order to solve the previous equations, we introduce a $2\times 2$ matrix describing the coupling between the pairs of equations due to the  rates $w_{i\to j}$ of the transitions between the reactive states
\be
{\boldsymbol{\mathsf W}}_0 \equiv \left(
\begin{array}{cc}
-w_{1\to 2} & w_{2\to 1}\\
w_{1\to 2} & -w_{2\to 1}
\end{array}
\right) ,
\label{W0-dfn}
\ee
as well as other $2\times 2$ matrices with the rates of the attachment and detachment reactions
\be
{\boldsymbol{\mathsf W}}_{\pm m} \equiv \left(
\begin{array}{cc}
w_{1,\pm m} & 0\\
0 & w_{2,\pm m}
\end{array}
\right) ,
\ee
together with
\be
{\boldsymbol{\mathsf A}} \equiv \sum_m {\boldsymbol{\mathsf W}}_{+m}\, .
\ee
Furthermore, we also define the $2\times 2$ matrices
\bea
\pmb{\Psi} &\equiv& \left(
\begin{array}{cc}
\psi(1) & \psi(1)\\
\psi(2) & \psi(2)
\end{array} \right) , \label{Psi-dfn}\\
\pmb{\Psi}(m_l) &\equiv& \left(
\begin{array}{cc}
\psi(m_l,1) & \psi(m_l,1)\\
\psi(m_l,2) & \psi(m_l,2)
\end{array} \right) , \\
\pmb{\Psi}(m_{l-1}m_l) &\equiv& \left(
\begin{array}{cc}
\psi(m_{l-1}m_l,1) & \psi(m_{l-1}m_l,1)\\
\psi(m_{l-1}m_l,2) & \psi(m_{l-1}m_l,2)
\end{array} \right) , \quad \\
&\vdots& \nonumber
\eea
As before, we have that
\bea
\sum_{m_l} \pmb{\Psi}(m_l)  &=& \pmb{\Psi} \, , \label{sum-Psi-1}\\
\sum_{m_{l-1}} \pmb{\Psi}(m_{l-1}m_l)  &=& \pmb{\Psi}(m_l) \, , \label{sum-Psi-2}\\
&\vdots& \nonumber
\eea
With these definitions, Eqs.~(\ref{eq-psi0-i}), (\ref{eq-psi1-i}),... of the hierarchy can be rewritten in the following matricial form:
\bea
0 &=& {\boldsymbol{\mathsf W}}_0 \cdot\pmb{\Psi} \, , \label{eq-Psi-0}\\
0 &=& {\boldsymbol{\mathsf W}}_{+m_l} \cdot\pmb{\Psi} + \sum_{m_{l+1}} {\boldsymbol{\mathsf W}}_{-m_{l+1}} \cdot \pmb{\Psi}(m_lm_{l+1}) \nonumber\\
&&\ \ - \left({\boldsymbol{\mathsf A}} + {\boldsymbol{\mathsf W}}_{-m_l} - {\boldsymbol{\mathsf W}}_0\right)\cdot\pmb{\Psi}(m_l) \, , \label{eq-Psi-1}\\
0 &=& {\boldsymbol{\mathsf W}}_{+m_l} \cdot\pmb{\Psi}(m_{l-1}) + \sum_{m_{l+1}} {\boldsymbol{\mathsf W}}_{-m_{l+1}} \cdot \pmb{\Psi}(m_{l-1}m_lm_{l+1})\nonumber\\
&&\ \  - \left({\boldsymbol{\mathsf A}} + {\boldsymbol{\mathsf W}}_{-m_l} - {\boldsymbol{\mathsf W}}_0\right)\cdot\pmb{\Psi}(m_{l-1}m_l) \, , \label{eq-Psi-2}\\
&& \qquad\qquad\qquad\qquad\qquad\vdots \nonumber
\eea

First, we note that Eqs.~(\ref{eq-psi0-i}) and thus~(\ref{eq-Psi-0}) can be solved with
\bea
&& \psi(1) = \frac{w_{2\to 1}}{w_{1\to 2} + w_{2\to 1}} \, , \\
&& \psi(2) = \frac{w_{1\to 2}}{w_{1\to 2} + w_{2\to 1}} \, , 
\eea
satisfying the normalization condition~(\ref{norm}).

Next, we assume that the following matricial factorization holds,
\be
\pmb{\Psi}(m_{l-r+1}\cdots m_{l-1}m_l) = {\boldsymbol{\mathsf Y}}_{m_l} \cdot {\boldsymbol{\mathsf Y}}_{m_{l-1}} \cdots {\boldsymbol{\mathsf Y}}_{m_{l-r+1}}\cdot\pmb{\Psi} 
\label{ansatz}
\ee
in terms of $2\times 2$ matrices ${\boldsymbol{\mathsf Y}}_{m}$ to be determined.  This factorization is suggested by the results of Coleman and Fox\cite{CF63JCP} in the fully irreversible regime.
Now, substituting the assumption~(\ref{ansatz}) into the equations~(\ref{eq-Psi-1}), (\ref{eq-Psi-2}),... of the hierarchy, we observe that they can all be solved if the matrices ${\boldsymbol{\mathsf Y}}_{m}$ satisfy the following relation,
\be
0 = {\boldsymbol{\mathsf W}}_{+m} + \sum_{m'} {\boldsymbol{\mathsf W}}_{-m'} \cdot {\boldsymbol{\mathsf Y}}_{m'}\cdot {\boldsymbol{\mathsf Y}}_{m} - \left({\boldsymbol{\mathsf A}} + {\boldsymbol{\mathsf W}}_{-m} - {\boldsymbol{\mathsf W}}_0\right)\cdot{\boldsymbol{\mathsf Y}}_{m} \, . \label{eq-X}
\ee
Introducing the $2\times 2$ matrix
\be
{\boldsymbol{\mathsf V}}\equiv {\boldsymbol{\mathsf A}}-\sum_{m} {\boldsymbol{\mathsf W}}_{-m} \cdot {\boldsymbol{\mathsf Y}}_{m} = \sum_{m} \left({\boldsymbol{\mathsf W}}_{+m}-{\boldsymbol{\mathsf W}}_{-m} \cdot {\boldsymbol{\mathsf Y}}_{m}\right) \, ,
\label{V-dfn}
\ee
we see that the solution of Eq.~(\ref{eq-X}) can be written in the form
\be
{\boldsymbol{\mathsf Y}}_{m} = ({\boldsymbol{\mathsf V}}-{\boldsymbol{\mathsf W}}_0+{\boldsymbol{\mathsf W}}_{-m})^{-1}\cdot {\boldsymbol{\mathsf W}}_{+m} \, .  
\label{eq-X-V}
\ee
Inserting Eq.~(\ref{eq-X-V}) back into Eq.~(\ref{V-dfn}), we obtain the self-consistent matrix equation
\be
\boxed{
{\boldsymbol{\mathsf V}}= ({\boldsymbol{\mathsf V}}-{\boldsymbol{\mathsf W}}_0)\cdot \sum_m ({\boldsymbol{\mathsf V}}-{\boldsymbol{\mathsf W}}_0+ {\boldsymbol{\mathsf W}}_{-m})^{-1} \cdot {\boldsymbol{\mathsf W}}_{+m} 
}
\label{eq-V}
\ee
that can be solved by direct numerical iteration for the given values of the rates. Finally, the mean growth velocity~(\ref{v}) is found to be equal to
\be
\boxed{v = {\rm tr} ({\boldsymbol{\mathsf V}}\cdot\pmb{\Psi})}
\label{v-V-Psi}
\ee
with ${\rm tr}$ denoting the trace of $2\times 2$ matrices.  In this regard, the matrix~(\ref{V-dfn}) is called the velocity matrix.  Once this latter is obtained by solving the self-consistent equation~(\ref{eq-V}), we can get the matrices~(\ref{eq-X-V}), which determine the composition of the copolymer sequences.

Besides, the diffusivity $\cal D$ can be calculated similarly at next order $q^2$.  According to the central limit theorem, the length probability distribution will thus be given in the long-time limit by the following Gaussian distribution
\be
p_t(l) \simeq \frac{1}{\sqrt{4\pi{\cal D} t}} \, \exp\left[ - \frac{(l-vt)^2}{4{\cal D}t} \right] .
\label{CLT}
\ee
Therefore, the solution of the coupled kinetic equations can be expressed for $t\to\infty$ as
\be
P_t(m_1\cdots m_l,i) \simeq p_t(l) \, \psi(m_1\cdots m_l,i)
\label{proba-t}
\ee
in terms of Eq.~(\ref{CLT}) and the stationary probability distribution
\be
\psi(m_1\cdots m_l,i) = \sum_{i_1\cdots i_l} \left({\boldsymbol{\mathsf Y}}_{m_l}\right)_{ii_l}  \cdots \left({\boldsymbol{\mathsf Y}}_{m_1}\right)_{i_2i_1}\left(\pmb{\Psi}\right)_{i_1i}
\label{proba}
\ee
 to find the sequence $m_1\cdots m_l$ of length $l$ in the reactive state $i$.  The normalization condition
\be
\sum_{m_1\cdots m_l,i} \psi(m_1\cdots m_l,i) = 1 
\ee
is satisfied because of Eqs.~(\ref{norm}), (\ref{sum-Psi-1}), (\ref{sum-Psi-2}),...
Moreover, the stationary probability to find the sequence $m_1\cdots m_l$ of length $l$ in any reactive state is given by the expression
\bea
&&\mu(m_1\cdots m_l) = \sum_i \psi(m_1\cdots m_l,i) \nonumber\\
&&= {\rm tr}\, \pmb{\Psi}(m_1\cdots m_l) = {\rm tr}\left(\,{\boldsymbol{\mathsf Y}}_{m_l}  \cdots {\boldsymbol{\mathsf Y}}_{m_1}\cdot\pmb{\Psi}\right) .
\label{mu}
\eea
Since these probabilities cannot be factorized into conditional probabilities as for Markov chains, we conclude that the sequences are nonMarkovian.

All these results can be extended to mechanisms with more than two reactive states, replacing the $2\times 2$ matrices by corresponding $I\times I$ matrices where $I$ is the number of reactive states.

We note that the results of Ref.~\onlinecite{AG09} are recovered if $w_{i\to j}=0$, as shown in Appendix~\ref{AppA}.

\subsection{Mean, variance, and correlation function}

In order to satisfy Eqs.~(\ref{sum-Psi-1}), (\ref{sum-Psi-2}),..., we have the property that
\be
{\boldsymbol{\mathsf R}}\cdot\pmb{\Psi}=\pmb{\Psi} \qquad\mbox{for}\qquad {\boldsymbol{\mathsf R}} \equiv \sum_m {\boldsymbol{\mathsf Y}}_{m} \, ,
\label{R}
\ee
implying 
\be
{\boldsymbol{\mathsf R}}^n\cdot\pmb{\Psi}=\pmb{\Psi} \qquad\forall \, n \in {\mathbb N} \, .
\ee
Because of the definition~(\ref{Psi-dfn}), the vector
\be
\pmb{\xi}_1 \equiv \left(
\begin{array}{c}
\psi(1) \\
\psi(2)
\end{array}
\right)
\label{xi-1}
\ee
is the right-eigenvector of the $2\times 2$ matrix ${\boldsymbol{\mathsf R}}$ associated with the eigenvalue $\Lambda_1=1$.  The corresponding left-eigenvector is denoted by the vector $\pmb{\eta}_1$.  We thus have that
\be
{\boldsymbol{\mathsf R}} =\pmb{\xi}_1 \, \Lambda_1 \, \pmb{\eta}_1^{\rm T} + \pmb{\xi}_2 \, \Lambda_2 \, \pmb{\eta}_2^{\rm T}
\label{S-decomp}
\ee
with a second eigenvalue $\vert\Lambda_2\vert <1$ and the biorthonormality condition $\pmb{\eta}_{\alpha}^{\rm T}\cdot\pmb{\xi}_{\beta}=\delta_{\alpha\beta}$, where the superscript ${\rm T}$ denotes the transpose.

The mean value of some function $f(m)$ of the monomeric unit $m$ at the location $k$ of the chain is defined by
\be
\langle f(m_k)\rangle = \sum_{m_1\cdots m_k \cdots m_l} \, f(m_k) \, \mu(m_1\cdots m_k \cdots m_l) \, .
\ee
Using Eqs.~(\ref{mu}) and~(\ref{R}), we get
\be
\langle f(m_k)\rangle =  {\rm tr}\left(\,{\boldsymbol{\mathsf R}}^{l-k}  \cdot {\boldsymbol{\mathsf F}} \cdot{\boldsymbol{\mathsf R}}^{k-1}\cdot\pmb{\Psi}\right) =  {\rm tr}\left(\,{\boldsymbol{\mathsf R}}^{l-k}  \cdot {\boldsymbol{\mathsf F}}\cdot\pmb{\Psi}\right)
\ee
where
\be
{\boldsymbol{\mathsf F}} \equiv \sum_m f(m) \, {\boldsymbol{\mathsf Y}}_{m} \, ,
\ee
so that the mean value is given by
\be
\langle f \rangle = \lim_{l-k\to\infty}\langle f(m_k)\rangle = \pmb{\eta}_{1}^{\rm T}\cdot {\boldsymbol{\mathsf F}}\cdot\pmb{\xi}_{1} \, .
\ee

The variance ${\rm Var}(f)\equiv \langle f^2\rangle - \langle f\rangle^2$ can be evaluated similarly.

Besides, the correlation function of the function $f$ is defined by
\be
\Gamma(n)\equiv \lim_{l-k\to \infty} \left[\langle f(m_k)f(m_{k+n})\rangle - \langle f\rangle^2\right] ,
\ee
where 
\bea
&& \langle f(m_k)f(m_{k+n})\rangle= \sum_{m_1\cdots m_k \cdots m_{k+n}\cdots m_l} \, f(m_k) \, f(m_{k+n}) \quad\nonumber\\
&&\qquad\qquad\qquad\qquad\ \times \mu(m_1\cdots m_k \cdots m_{k+n} \cdots m_l)\, ,
\eea
giving $\langle f^2 \rangle$ for $n=0$, and
\be
\langle f(m_k)\, f(m_{k+n})\rangle =  {\rm tr}\left(\,{\boldsymbol{\mathsf R}}^{l-k-n}  \cdot {\boldsymbol{\mathsf F}} \cdot{\boldsymbol{\mathsf R}}^{n-1}\cdot {\boldsymbol{\mathsf F}} \cdot\pmb{\Psi}\right)
\ee
for $n>0$. Consequently, we have that
\bea
&&\lim_{l-k\to\infty}\langle f(m_k)\, f(m_{k+n})\rangle = \left(\pmb{\eta}_{1}^{\rm T}\cdot {\boldsymbol{\mathsf F}}\cdot\pmb{\xi}_{1}\right)^2 \quad\nonumber\\
&&\qquad\qquad + \Lambda_2^{n-1} \left(\pmb{\eta}_{1}^{\rm T}\cdot {\boldsymbol{\mathsf F}}\cdot\pmb{\xi}_{2}\right)\left(\pmb{\eta}_{2}^{\rm T}\cdot {\boldsymbol{\mathsf F}}\cdot\pmb{\xi}_{1}\right)  .
\eea
Therefore, the correlation function is given by $\Gamma(0)={\rm Var}(f)$ for $n=0$, and
\be
\Gamma(n) =  \Lambda_2^{n-1} \left(\pmb{\eta}_{1}^{\rm T}\cdot {\boldsymbol{\mathsf F}}\cdot\pmb{\xi}_{2}\right)\left(\pmb{\eta}_{2}^{\rm T}\cdot {\boldsymbol{\mathsf F}}\cdot\pmb{\xi}_{1}\right)
\label{correl}
\ee
for $n>0$. Since $\vert \Lambda_2\vert <1$,  the correlation function decreases exponentially as $\Gamma(n)\sim\exp(-\gamma n)$ for $n\to\infty$ with the rate $\gamma=-\ln\vert \Lambda_2\vert$.  In the following, we shall use the normalized correlation function, $C(n)\equiv \Gamma(n)/\Gamma(0)$.

\subsection{Bulk probabilities}

The bulk probability $\bar{\mu}(m)$ to find the monomeric unit $m$ anywhere inside the grown copolymer chain can be obtained as the mean value of the indicator function of the monomeric unit $m$ given by the corresponding Kronecker symbol: $f(m_k)=\sigma_m(m_k)=\delta_{m,m_k}$.   Therefore, we get
\be
\bar{\mu}(m)=\langle \sigma_m\rangle = \pmb{\eta}_{1}^{\rm T}\cdot {\boldsymbol{\mathsf Y}}_{m}\cdot\pmb{\xi}_{1} \, .
\label{bulk-proba-1}
\ee
Similarly, the bulk probability to find the subsequence $m_1\cdots m_k$ anywhere in the chain is given by 
\be
\bar{\mu}(m_1\cdots m_k)=\langle \sigma_{m_1}\cdots\sigma_{m_k}\rangle = \pmb{\eta}_{1}^{\rm T}\cdot {\boldsymbol{\mathsf Y}}_{m_k}\cdots{\boldsymbol{\mathsf Y}}_{m_1}\cdot\pmb{\xi}_{1} \, .
\label{bulk-proba-k}
\ee
If we introduce the $2\times 2$ matrix
\be
\pmb{\Upsilon}\equiv \pmb{\xi}_{1}\,\pmb{\eta}_{1}^{\rm T}\, ,
\label{Upsilon}
\ee 
Eq.~(\ref{bulk-proba-k}) reads
\be
\bar{\mu}(m_1\cdots m_k)={\rm tr}\left({\boldsymbol{\mathsf Y}}_{m_k}\cdots{\boldsymbol{\mathsf Y}}_{m_1}\cdot\pmb{\Upsilon}\right) .
\label{bulk-proba-k-bis}
\ee
In general, neither the tip probabilities~(\ref{mu}), nor the bulk probabilities~(\ref{bulk-proba-k-bis}) have an expression that factorizes as for Bernoulli or Markov chains, which confirms the nonMarkovian character of the macromolecular chains yielded by multistate copolymerization processes.

\subsection{Fully irreversible regime}

This regime has been studied in Ref.~\onlinecite{CF63JCP} in the context of polymerization with isotactic and syndiotactic placements, correponding respectively to $m={\rm I}$ and $m={\rm S}$.  Fully irreversible regimes are defined by supposing that the detachment rates are negligible in front of the attachment rates:
\be
{\boldsymbol{\mathsf W}}_{-m} = 0 \, .
\label{zero-detachment}
\ee
Accordingly, the solution of Eq.~(\ref{eq-V}) is given by
\be
{\boldsymbol{\mathsf V}}= \sum_m {\boldsymbol{\mathsf W}}_{+m} 
\label{eq-V-FI}
\ee
and the mean growth velocity~(\ref{v-V-Psi}) is thus equal to
\be
v = \frac{w_{2\to 1}}{w_{1\to 2} + w_{2\to 1}}\, a_1 + \frac{w_{1\to 2}}{w_{1\to 2} + w_{2\to 1}} \, a_2\, ,
\ee
which reads
\be
v = \frac{\lambda_a}{\lambda_a + \lambda_b}\, k_1 \, [{\rm M}] + \frac{\lambda_b}{\lambda_a + \lambda_b} \, k_2 \, [{\rm M}] \, ,
\label{v-CF63}
\ee
with the notations $k_i=\sum_m k_{i,+m}$ ($i=1,2$), $\lambda_a=w_{2\to 1}$, $\lambda_b=w_{1\to 2}$, and the monomeric concentration $c_m=[{\rm M}]$, showing the equivalence with Eqs. (2.4a)-(2.5) of Ref.~\onlinecite{CF63JCP}.
According to Eq.~(\ref{eq-X-V}), we also have that
\be
{\boldsymbol{\mathsf Y}}_{m} = ({\boldsymbol{\mathsf V}}-{\boldsymbol{\mathsf W}}_0)^{-1}\cdot {\boldsymbol{\mathsf W}}_{+m} \, .  
\label{eq-X-V-irr}
\ee
In order to compare with the related results of Ref.~\onlinecite{CF63JCP}, we introduce the transformation
\be
{\boldsymbol{\mathsf T}}  \equiv \left(
\begin{array}{cc}
w_{2\to 1} & 0\\
0 & w_{1\to 2} 
\end{array}
\right) = \left(
\begin{array}{cc}
\lambda_a & 0\\
0 & \lambda_b 
\end{array}
\right) ,
\label{T-dfn}
\ee
such that
\be
{\boldsymbol{\mathsf W}}_0={\boldsymbol{\mathsf T}}\cdot{\boldsymbol{\mathsf W}}_0^{\rm T} \cdot{\boldsymbol{\mathsf T}}^{-1} \, .
\ee
The correspondence with the matrices used in Ref.~\onlinecite{CF63JCP} is established according to
\bea
&&\pmb{\Phi} \equiv {\boldsymbol{\mathsf T}}\cdot\pmb{\Upsilon}^{\rm T} \cdot{\boldsymbol{\mathsf T}}^{-1} \, , \\
&& {\boldsymbol{\mathsf X}}_m={\boldsymbol{\mathsf T}}\cdot{\boldsymbol{\mathsf Y}}_m^{\rm T} \cdot{\boldsymbol{\mathsf T}}^{-1} \, ,
\eea
for $m\in\{{\rm I},{\rm S}\}$. A direct calculation using Mathematica\cite{Mathematica} shows that
\be
\pmb{\Phi} \equiv \left(
\begin{array}{cc}
\phi_1 & \phi_1\\
\phi_2 & \phi_2
\end{array} \right) ,
\ee
where
\be
\phi_1 = \frac{k_1 \lambda_a}{k_1 \lambda_a+k_2 \lambda_b} \, , \qquad
\phi_2 = \frac{k_2 \lambda_b}{k_1 \lambda_a+k_2 \lambda_b} \, ,
\ee
with $k_i=k_{i,+{\rm I}}+k_{i,+{\rm S}}$, which corresponds to Eq.~(2.7) of Ref.~\onlinecite{CF63JCP}, and
\bea
&&\left({\boldsymbol{\mathsf X}}_{m}\right)_{11}=\frac{k_{1,+{m}}(\lambda_a+k_2[{\rm M}])}{k_1\lambda_a+k_2\lambda_b+k_1 k_2 [{\rm M}]} \, , \\
&&\left({\boldsymbol{\mathsf X}}_{m}\right)_{22}=\frac{k_{2,+{m}}(\lambda_b+k_1[{\rm M}])}{k_1\lambda_a+k_2\lambda_b+k_1 k_2 [{\rm M}]} \, , \\
&&\left({\boldsymbol{\mathsf X}}_{m}\right)_{12}=\frac{k_{1,+{m}}\lambda_a}{k_1\lambda_a+k_2\lambda_b+k_1 k_2 [{\rm M}]} \, , \\
&&\left({\boldsymbol{\mathsf X}}_{m}\right)_{21}=\frac{k_{2,+{m}}\lambda_b}{k_1\lambda_a+k_2\lambda_b+k_1 k_2 [{\rm M}]} \, , 
\eea
for $m\in\{{\rm I},{\rm S}\}$, which corresponds to Eqs.~(3.9a)-(3.9d) of Ref.~\onlinecite{CF63JCP}.  Therefore, the results of Ref.~\onlinecite{CF63JCP} are precisely recovered in the fully irreversible regime.

\section{Thermodynamics}
\label{Thermo}

\subsection{Entropy production}

For isothermal-isobaric stochastic processes ruled by kinetic equations
\bea
\frac{d}{dt}\, P_t(\omega) &=& \sum_{\omega'(\neq\omega)} \big[P_t(\omega')\,  W(\omega'\to\omega) \nonumber\\
&&\qquad\qquad\qquad - P_t(\omega)\,  W(\omega\to\omega') \big] ,
\label{gen-master-eq}
\eea
where $W(\omega\to\omega')$ are the transition rates between the coarse-grained states $\omega$ and $\omega'$, the link to thermodynamics is established by using the relations
\be
\frac{W(\omega\to\omega')}{W(\omega'\to\omega)} = {\rm e}^{\beta\left[ G(\omega)-G(\omega')\right]}
\label{W-ratio}
\ee
giving the ratio of the rates of opposite transitions in terms of Gibbs' free energies $G(\omega)$ associated with the coarse-grained states $\omega$, while $\beta=(k_{\rm B}T)^{-1}$ is the inverse temperature expressed in terms of the temperature $T$ and Boltzmann's constant $k_{\rm B}$.  Gibbs' free energy $G(\omega)$ is related to the enthalpy $H(\omega)$ and the entropy $S(\omega)$ of the coarse-grained state $\omega$ by $G(\omega)=H(\omega)-TS(\omega)$.  Furthermore, the overall thermodynamic entropy of the statistical sample described by the probability distribution $P_t(\omega)$ is defined by
\be
S_t = \sum_{\omega} P_t(\omega) \left[ S(\omega) - k_{\rm B} \ln P_t(\omega)\right] .
\label{entropy}
\ee
In general, the time derivative $dS_t/dt$ of the entropy can be separated into the rate of entropy exchange with the environment of the growing copolymer
\be
\frac{d_{\rm e}S}{dt} = \frac{1}{T} \, \frac{d\langle H\rangle_t}{dt} 
\label{deSdt}
\ee
expressed in terms of the mean enthalpy $\langle H\rangle_t = \sum_{\omega} P_t(\omega)\, H(\omega)$, and the rate of entropy production
\bea
\frac{d_{\rm i}S}{dt} &=& \frac{dS_t}{dt} - \frac{d_{\rm e}S}{dt} \nonumber\\
&=& -\frac{1}{T} \, \frac{d\langle G\rangle_t}{dt} - k_{\rm B}\, \frac{d}{dt} \sum_{\omega} P_t(\omega) \, \ln P_t(\omega) \geq 0 , \quad
\label{diSdt}
\eea
which is always non-negative in accordance with the second law of thermodynamics. 

These considerations apply in particular to the kinetic equations~(\ref{eq-i}) where the coarse-grained state $\omega$ is defined by the sequence $m_1\cdots m_l$ in the reactive state~$i$. 

In the regime of steady growth where the mean growth velocity is positive $v>0$, the entropy production rate~(\ref{diSdt}) can be evaluated using the probability distribution~(\ref{proba-t}) that is the solution of the kinetic equations in the long-time limit.\cite{AG09,AG08}  On the one hand, the mean value of the Gibbs free energy is given by $\langle G\rangle_t \simeq \langle l\rangle_t g$ in terms of the mean length $\langle l \rangle_t\simeq vt$, so that
\be
\frac{d\langle G\rangle_t}{dt} = v \, g
\label{g-v}
\ee
with the mean Gibbs free energy per monomeric unit
\be
g \equiv \lim_{l\to\infty} \frac{1}{l} \sum_{m_1\cdots m_l, i} \psi(m_1\cdots m_l, i) \, G(m_1\cdots m_l, i) \, .
\label{g}
\ee
On the other hand, Eq.~(\ref{proba-t}) also implies that
\be
-\frac{d}{dt}\sum_{\omega} P_t(\omega) \, \ln P_t(\omega) = v \, D
\ee
with the sequence disorder per monomeric unit
\be
D \equiv \lim_{l\to\infty} - \frac{1}{l} \sum_{m_1\cdots m_l, i} \psi(m_1\cdots m_l, i) \ln \psi(m_1\cdots m_l, i) , \qquad
\label{D-dfn}
\ee
which is always a non-negative quantity, $D\geq 0$.
Therefore, the entropy production rate can be expressed as
\be
\frac{d_{\rm i}S}{dt} = k_{\rm B} \, v \, A \geq 0 \, ,
\label{epr-copolym}
\ee
with the dimensionless entropy production per monomeric unit, called the affinity,
\be
A \equiv \varepsilon + D \, ,
\label{Aff}
\ee
the dimensionless free-energy driving force defined in terms of Eq.~(\ref{g}) and the thermal energy $k_{\rm B}T$ as
\be
\varepsilon \equiv - \frac{g}{k_{\rm B}T} ,
\ee 
and the sequence disorder defined by Eq.~(\ref{D-dfn}).\cite{AG09,AG08}

Computing the time derivative of the mean Gibbs free energy $\langle G\rangle_t$ with the kinetic equations~(\ref{gen-master-eq}), we get
\bea
&&\frac{d\langle G\rangle_t}{dt} = \sum_{\omega\neq\omega'} P_t(\omega)\,  W(\omega\to\omega') \left[ G(\omega')-G(\omega)\right] \nonumber\\
&&= -k_{\rm B}T \sum_{\omega\neq\omega'} P_t(\omega)\,  W(\omega\to\omega') \, \ln \frac{W(\omega\to\omega')}{W(\omega'\to\omega)} ,
\eea
where we used Eq.~(\ref{W-ratio}).  According to Eqs.~(\ref{proba-t}) and~(\ref{g-v}), we find that the free-energy driving force is here given by
\be
\varepsilon = \frac{1}{v} \sum_{m,i} \left[\psi(i) \, w_{i,+m} - \psi(m,i) \, w_{i,-m}\right] \, \ln \frac{w_{i,+m}}{w_{i,-m}} \, .
\label{eps}
\ee
We notice that the velocity is similarly expressed by Eq.~(\ref{v}).

If the detachment rates are negligible in front of the attachment rates, the free-energy driving force~(\ref{eps}) is arbitrarily large, so that the entropy production rate becomes infinite in the fully irreversible regime.

\subsection{The equilibrium limit}

The state of thermodynamic equilibrium is identified by the principle of detailed balance, which requires that
\bea
&& w_{1\to 2} \, P_{\rm eq}(m_1\cdots m_l, 1) = w_{2\to 1} \, P_{\rm eq}(m_1\cdots m_l, 2) , \quad \label{det-bal-12}\\
&& w_{1,+m_l} \, P_{\rm eq}(m_1\cdots m_{l-1}, 1) \nonumber\\
&&\qquad\qquad\qquad = w_{1,-m_l} \, P_{\rm eq}(m_1\cdots m_{l-1}m_l, 1) , \label{det-bal-1}\\
&& w_{2,+m_l} \, P_{\rm eq}(m_1\cdots m_{l-1}, 2) \nonumber\\
&&\qquad\qquad\qquad  = w_{2,-m_l} \, P_{\rm eq}(m_1\cdots m_{l-1}m_l, 2) . \label{det-bal-2}
\eea
This imposes the following constraints on the attachment and detachment rates:
\be
\frac{w_{1,+m}}{w_{1,-m}} = \frac{w_{2,+m}}{w_{2,-m}}  
\ee
for all $m=1,2,...,M$.  If the attachment rates are proportional to the corresponding monomeric concentrations according to the mass action law, we thus have that
\be
\frac{w_{i,+m}}{w_{i,-m}} = \frac{k_{i,+m}}{k_{i,-m}}  \equiv {\rm e}^{\varepsilon_m}
\label{rates-equil}
\ee
for all $m=1,2,...,M$, independently of the reactive state~$i$.  These constraints must be satisfied for the existence of a thermodynamic equilibrium limit.

Now, taking $l=1$ and $m=m_l$ in Eqs.~(\ref{det-bal-1})-(\ref{det-bal-2}), and using Eq.~(\ref{rates-equil}),  we get
\be
\psi_{\rm eq}(m,i) = {\rm e}^{\varepsilon_m} \, \psi_{\rm eq}(i)
\ee
for $i=1,2$.  Therefore, the equilibrium tip probabilities are given by
\be
\mu_{\rm eq}(m) = \sum_i \psi_{\rm eq}(m,i) = {\rm e}^{\varepsilon_m}
\label{mu-eq}
\ee
because of Eq.~(\ref{norm}). Since the probability distribution~(\ref{mu-eq}) should also be normalized to unity, we must have
\be
\sum_{m=1}^M {\rm e}^{\varepsilon_m} = 1 \qquad \mbox{at equilibrium.}
\ee

Moreover, Eq.~(\ref{ansatz}) for $r=1$ gives $\psi(m,i)=\sum_j \left({\boldsymbol{\mathsf Y}}_{m}\right)_{ij} \psi(j)$, while the probabilities $\psi(i)$ define the vector~(\ref{xi-1}), so that
\be
{\boldsymbol{\mathsf Y}}_{m}\cdot\pmb{\xi}_1 =  {\rm e}^{\varepsilon_m} \, \pmb{\xi}_1 \qquad\mbox{at equilibrium.}
\label{equil-eigen}
\ee
As a consequence, the sequence probability~(\ref{mu}) factorizes as
\be
\mu_{\rm eq}(m_1\cdots m_l) = \prod_{k=1}^l \mu_{\rm eq}(m_k) \, ,
\label{Bernoulli}
\ee
showing that the sequences form a Bernoulli chain in the special limit where the equilibrium conditions~(\ref{det-bal-12})-(\ref{det-bal-2}) are satisfied.  Because of Eq.~(\ref{bulk-proba-1}), the equilibrium bulk probabilities of the monomeric units are also given by
\be
\bar{\mu}_{\rm eq}(m) =  {\rm e}^{\varepsilon_m} \, .
\label{bar-mu-eq}
\ee

Furthermore, the mean growth velocity is equal to zero in the equilibrium limit.  Indeed, Eq.~(\ref{rates-equil}) has the matrix form ${\boldsymbol{\mathsf W}}_{+m}={\rm e}^{\varepsilon_m}{\boldsymbol{\mathsf W}}_{-m}$, so that Eq.~(\ref{V-dfn}) implies that ${\boldsymbol{\mathsf V}}\cdot\pmb{\Psi}=0$ at equilibrium and the mean growth velocity~(\ref{v-V-Psi}) is thus vanishing.  Accordingly, we find that the equilibrium free-energy driving force~(\ref{eps}) can be expressed as
\be
\varepsilon_{\rm eq} = \sum_{m=1}^M \varepsilon_m \, \bar{\mu}_{\rm eq}(m)
\label{eps-eq}
\ee
in terms of the equilibrium bulk probabilities~(\ref{bar-mu-eq}).  The equilibrium value of the affinity~(\ref{Aff}) is also equal to zero, which implies that the equilibrium value of the sequence disorder is given by $D_{\rm eq}=-\varepsilon_{\rm eq}$, so that
\be
D_{\rm eq} = - \sum_{m=1}^{M} \bar{\mu}_{\rm eq}(m) \, \ln \bar{\mu}_{\rm eq}(m) \, ,
\label{D-eq}
\ee
in agreement with the result that the copolymer is a Bernoulli chain in the equilibrium limit.  As required, the entropy production rate is thus equal to zero at equilibrium.

Besides, the correlation function~(\ref{correl}) reduces to $\Gamma(n)={\rm Var}(f)\, \delta_{n,0}$ in the equilibrium limit.  Indeed, we have that $\pmb{\eta}_{2}^{\rm T}\cdot {\boldsymbol{\mathsf F}}\cdot\pmb{\xi}_{1} = \langle f\rangle_{\rm eq}\, \pmb{\eta}_{2}^{\rm T}\cdot \pmb{\xi}_{1}=0$ where $\langle f\rangle_{\rm eq}=\sum_m f(m) \mu_{\rm eq}(m)$, because of Eq.~(\ref{equil-eigen}) and the biorthonormality condition $\pmb{\eta}_{2}^{\rm T}\cdot \pmb{\xi}_{1}=0$ between the left- and right-eigenvectors associated with different eigenvalues.  There is thus no statistical correlation in the sequences, as expected since they form Bernoulli chains in the equilibrium limit.

\section{Illustrative examples}
\label{Examples}

In the following examples, $M=2$ monomeric species A for $m=1$ and B for $m=2$ are considered.  The theoretical predictions are calculated with the methods of Appendix~\ref{AppB} and the growth is simulated using Gillespie's algorithm,\cite{G76,G77} as described in Appendix~\ref{AppC}.

\subsection{Example with thermodynamic equilibrium}

The parameters of this example are taken as
\bea
&& k_{1,+{\rm A}}=2 \, , \quad k_{1,+{\rm B}}=4 \, , \quad k_{2,+{\rm A}}=4 \, , \quad k_{2,+{\rm B}}=2 \, , \nonumber\\
&& k_{1,-{\rm A}}=1 \, , \quad k_{1,-{\rm B}}=6 \, , \quad k_{2,-{\rm A}}=2 \, , \quad k_{2,-{\rm B}}=3 \, , \nonumber\\
&& k_{12}=1 \, , \qquad k_{21} = 2 \, , \qquad c_{\rm B}=1 \, .
\label{Ex2}
\eea
Here, the concentration $c_{\rm A}$ is the nonequilibrium control parameter.
This set of parameter values satisfies the condition~(\ref{rates-equil}) for the existence of an equilibrium limit at $c_{\rm A}=1/6$.  Accordingly, the thermodynamic equilibrium limit is reached at the concentration where the mean growth velocity vanishes.  

\begin{table}
\caption{\label{tab.Ex2} Growth of a copolymer in the conditions (\ref{Ex2}) with $c_{\rm A}=5$: Comparison between the values of the tip probabilities~(\ref{proba}) obtained by theory with $10^4$ iterations of Eq.~(\ref{eq-V}) and numerically by Gillespie's algorithm generating $10^7$ copolymer sequences, each of total time $t=100$.  Under the chosen conditions, the mean growth velocity is $v=14.580$, the free-energy driving force $\varepsilon = 1.8623$, the sequence disorder $D=0.4534$, and the affinity $A=\varepsilon+D=2.3157$.  The mean length of the sequences is thus equal to $\langle l\rangle_t=vt=1458$. The same simulation data are used for Fig.~\ref{fig1}. The numbers are rounded off to five decimal digits.}
\vspace{5mm}
\begin{center}
\begin{tabular}{|cccccc|}
\hline
$m_{l-2}$  & $m_{l-1}$ & $m_l\ \,$ & $\quad i \quad$ & theory & simulation \\
\hline
 & & & $1$ & $0.66667$ & $0.66671$  \\
  & & & $2$ & $0.33333$ & $0.33329$ \\
 & & A & $1$ & $0.52263$ & $0.52262$  \\
 & & B & $1$ & $0.14403$ & $0.14409$  \\
 & & A & $2$ & $0.30048$ & $0.30048$  \\
 & & B & $2$ & $0.03285$ & $0.03281$  \\
 & A & A & $1$ & $0.41483$ & $0.41494$  \\
 & B & A & $1$ & $0.10781$ & $0.10768$  \\
 & A & B & $1$ & $0.11328$ & $0.11331$  \\
 & B & B & $1$ & $0.03076$ & $ 0.03078$  \\
& A & A & $2$ & $0.26832$ & $0.26832$  \\
 & B & A & $2$ & $0.03217$ & $0.03216$  \\
 & A & B & $2$ & $0.02889$ & $0.02888$  \\
 & B & B & $2$ & $0.00396$ & $0.00393$  \\
A & A & A & $1$ & $0.33315$ & $0.33337$  \\
B & A & A & $1$ & $0.08168$ & $0.08158$  \\
A & B & A & $1$ & $0.08518$ & $0.08512$  \\
B & B & A & $1$ & $0.02262$ & $0.02257$  \\
A & A & B & $1$ & $ 0.09018$ & $0.09020$  \\
B & A & B & $1$ & $0.02309$ & $0.02311$  \\
A & B & B & $1$ & $0.02422$ & $0.02422$  \\
B & B & B & $1$ & $0.00654$ & $0.00655$  \\
A & A & A & $2$ & $0.23791$ & $0.23788$  \\
B & A & A & $2$ & $0.03041$ & $0.03044$  \\
A & B & A & $2$ & $0.02785$ & $0.02786$  \\
B & B & A & $2$ & $0.00431$ & $0.00430$  \\
A & A & B & $2$ & $0.02532$ & $0.02534$  \\
B & A & B & $2$ & $0.00357$ & $0.00355$  \\
A & B & B & $2$ & $0.00336$ & $0.00333$  \\
B & B & B & $2$ & $0.00060$ & $0.00059$  \\
\hline
\end{tabular}
\end{center}
\end{table} 

\begin{figure}[h]
\centerline{\scalebox{0.495}{\includegraphics{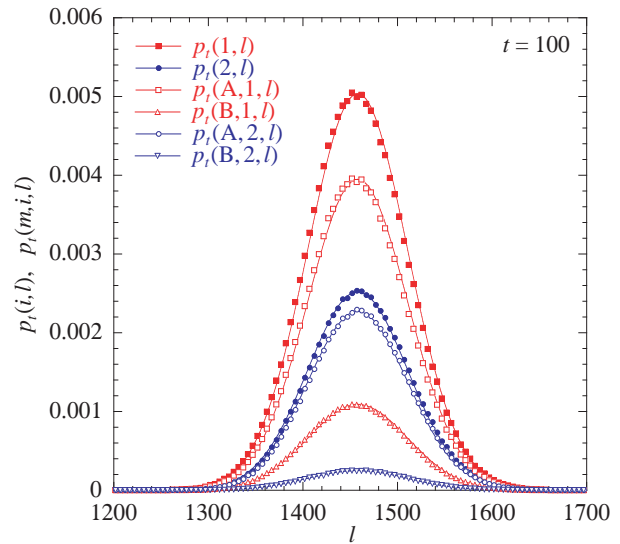}}}
\caption{Growth of a copolymer in the conditions (\ref{Ex2}) with $c_{\rm A}=5$: Probability distributions $p_t(i,l)$ and $p_t(m,i,l)$ at the time $t=100$, versus the length $l$.  The data points are obtained with a statistics of $10^7$ sequences generated by Gillespie's algorithm running over the total time $t=100$.  The lines are the theoretical predictions of Eqs.~(\ref{CLT})-(\ref{proba-t}) with the mean growth velocity $v=14.58$ and the diffusivity ${\cal D}=13.95$.}
\label{fig1}
\end{figure}

\begin{figure}[h]
\centerline{\scalebox{0.495}{\includegraphics{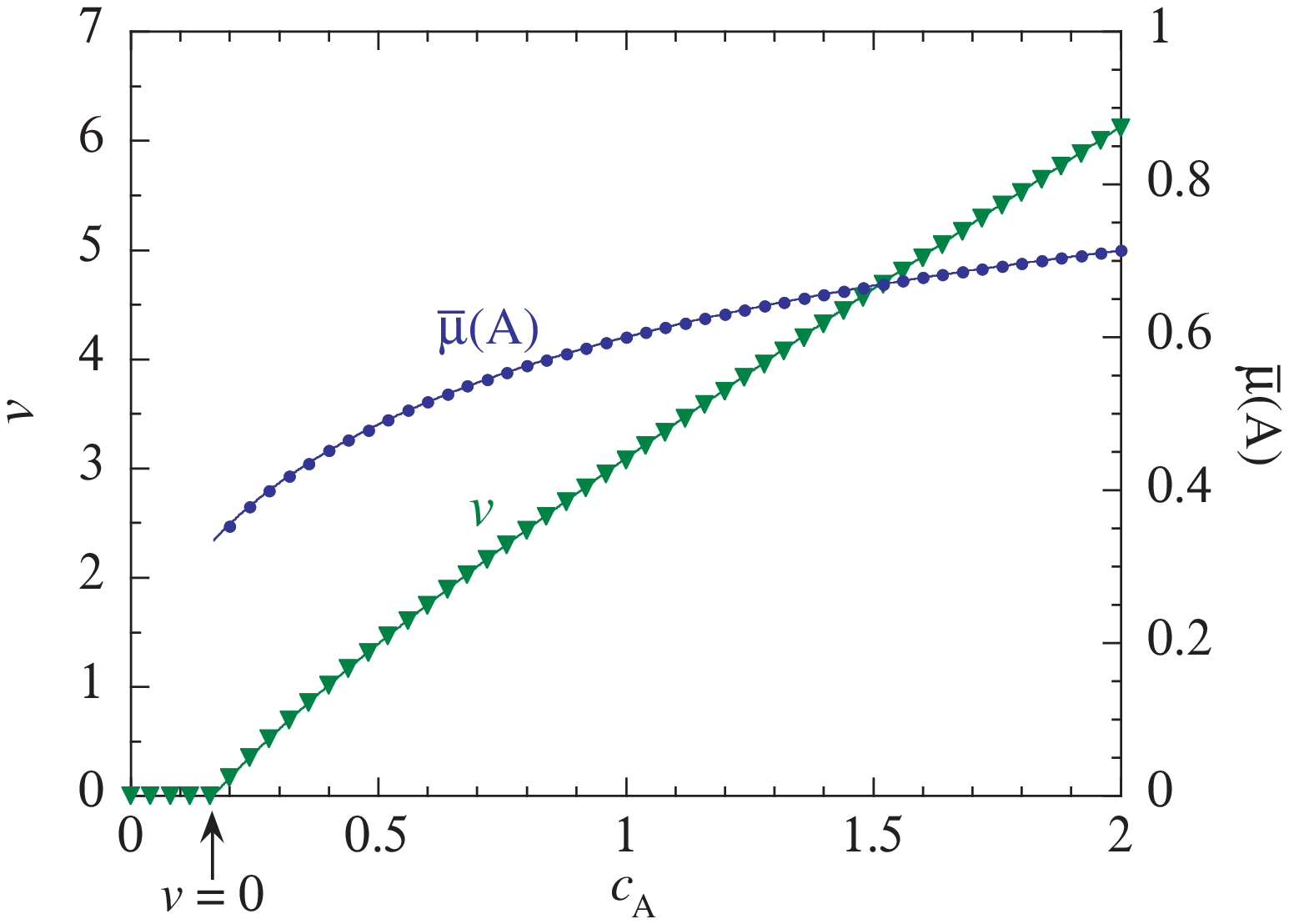}}}
\caption{Growth of a copolymer in the conditions (\ref{Ex2}): The mean growth velocity $v$ and the bulk probability $\bar{\mu}({\rm A})$ of the monomeric unit A, versus the concentrations $c_{\rm A}$.  The data points show the results of simulations with Gillespie's algorithm generating a sequence after $10^6$ jumps of the algorithm for every value of the concentration $c_{\rm A}$. The lines depict the theoretical predictions of Eqs.~(\ref{v-V-Psi}) and~(\ref{bulk-proba-1}).}
\label{fig2}
\end{figure}

\begin{figure}[h]
\centerline{\scalebox{0.495}{\includegraphics{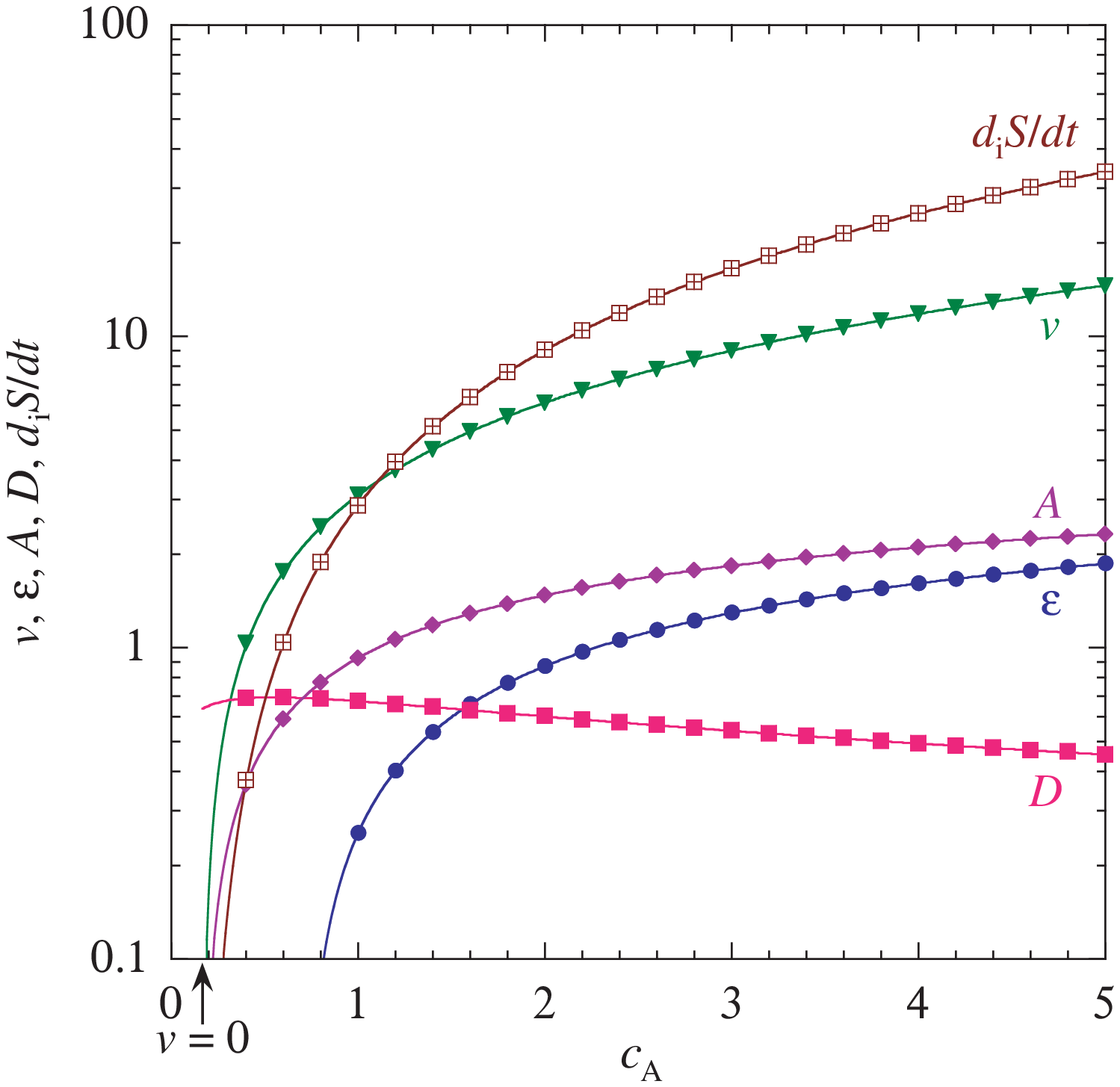}}}
\caption{Growth of a copolymer in the conditions (\ref{Ex2}): The mean growth velocity $v$, the free-energy driving force $\varepsilon$, the sequence disorder $D$, the affinity $A$, and the entropy production rate $d_{\rm i}S/dt$, versus the concentrations $c_{\rm A}$.  The data points are obtained with a statistics of $10^6$ sequences generated by Gillespie's algorithm running over the total time $t=100$.  
The lines show the theoretical predictions.}
\label{fig3}
\end{figure}

\begin{figure}[h]
\centerline{\scalebox{0.495}{\includegraphics{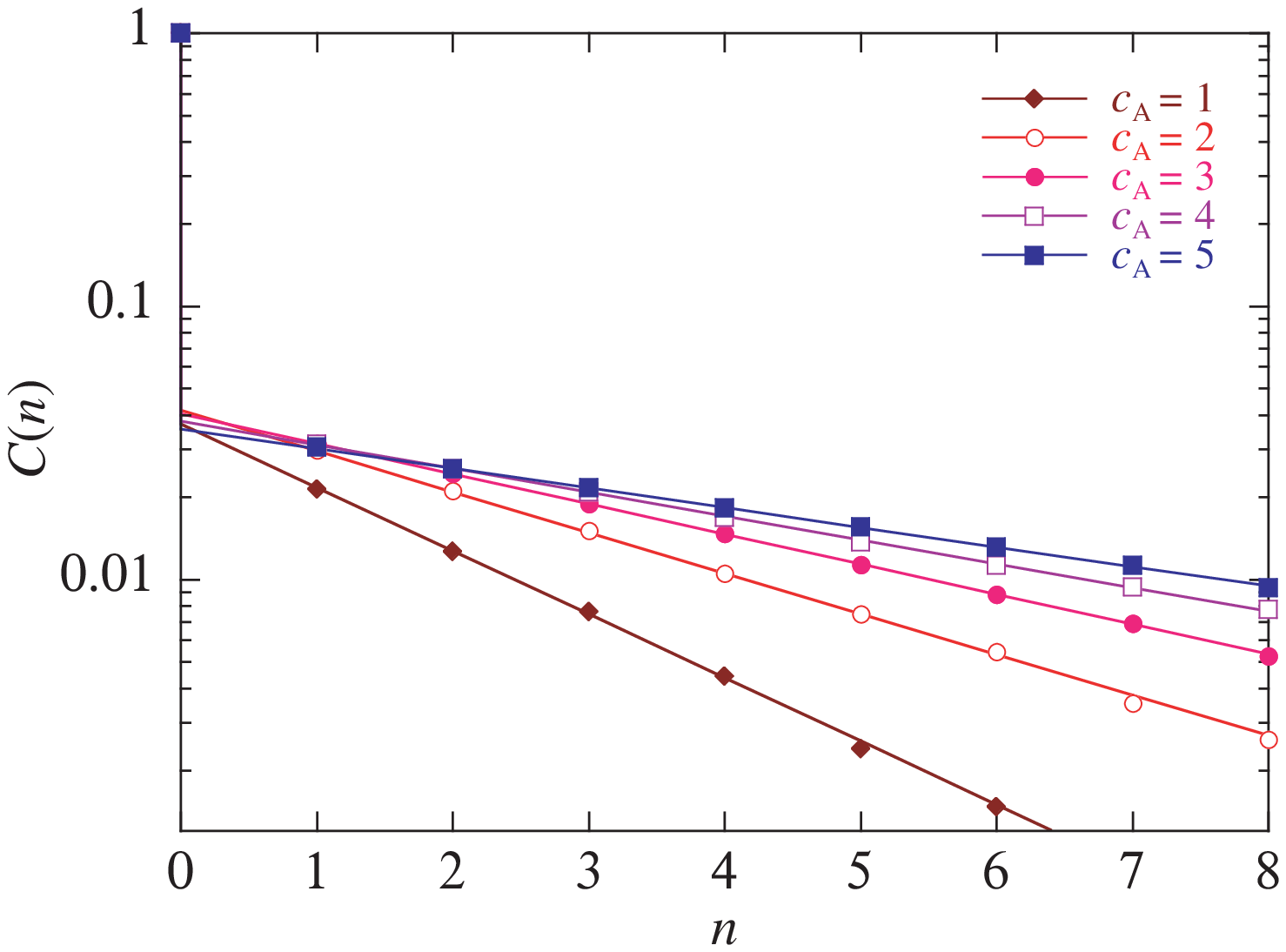}}}
\caption{Growth of a copolymer in the conditions (\ref{Ex2}):  The correlation function $C(n)$ versus $n$ for different values of the concentration $c_{\rm A}$.  The data points show the simulation results of Eq.~(\ref{correl-app}) using a long sequence generated by $10^8$ jumps of Gillespie's algorithm.  The total length of the sequence ranges from $L\simeq 3\times 10^7$ for the lowest concentration value $c_{\rm A}=1$ to $L\simeq 7.3\times 10^7$ for the largest one $c_{\rm A}=5$. The lines depict the theoretical predictions of Eq.~(\ref{correl}).}
\label{fig4}
\end{figure}

For this example, the tip probabilities~(\ref{proba}) have been computed using theory and simulations.  The results are compared at the concentration value $c_{\rm A}=5$ in Table~\ref{tab.Ex2}, showing the tip probabilities $\psi(i)$, $\psi(m_l,i)$, $\psi(m_{l-1}m_l,i)$, and $\psi(m_{l-2}m_{l-1}m_l,i)$.  On the one hand, these probabilities are obtained in theory by solving Eq.~(\ref{eq-V}) by iterations to get the velocity matrix that is next used to find the matrices~(\ref{eq-X-V}) and thus the tip probabilities~(\ref{proba}). On the other hand, the growth is simulated during the time lapse $t=100$ with Gillespie's algorithm to generate a sample of $10^7$ sequences.  With this sample, the probabilities $p_t(i,l)$, $p_t(m_l,i,l)$, $p_t(m_{l-1}m_l,i,l)$, and $p_t(m_{l-2}m_{l-1}m_l,i,l)$ are first computed by statistics to next obtain the tip probabilities by summing over the different values of the length $l$ to get $\psi(i)=\sum_l p_t(i,l)$, $\psi(m_l,i)=\sum_l p_t(m_l,i,l)$, etc...  In Table~\ref{tab.Ex2}, we see the excellent agreement between theory and simulations in this example.  The nonMarkovian character of these probability distribution is also confirmed by these results.  For the same data, Fig.~\ref{fig1} shows the probabilities $p_t(i,l)$ and $p_t(m_l,i,l)$ as a function of the length $l$ in comparison with the theoretical prediction~(\ref{CLT})-(\ref{proba-t}) that the probability distributions should be Gaussian after a long enough time according to the central limit theorem.  Here also, agreement is observed between theory (lines) and simulations (data points).

Furthermore, the different quantities of interest have been investigated as a function of the concentration $c_{\rm A}$.
Figure~\ref{fig2} depicts the mean growth velocity~(\ref{v-V-Psi}) and the bulk probability~(\ref{bulk-proba-1}) to find the monomeric unit A anywhere inside the grown chain versus the concentration~$c_{\rm A}$.  Again, the data points show the simulation results with Gillespie's algorithm and the lines the theoretical predictions.  As the concentration~$c_{\rm A}$ increases, the composition of the copolymer in monomeric units A also increases, as seen in Fig.~\ref{fig2}. 

The thermodynamic quantities are shown as a function of the concentration $c_{\rm A}$ in Fig.~\ref{fig3}.  In simulations, the free-energy driving force is computed by adding together the contributions of every jump, as explained in Appendix~\ref{AppC}.  In theory, it is given by Eq.~(\ref{eps}).  The sequence disorder is obtained with Eqs.~(\ref{D-l-dfn})-(\ref{D-diff}).  Next, the affinity is calculated by Eq.~(\ref{Aff}) and the entropy production rate by Eq.~(\ref{epr-copolym}) in units where $k_{\rm B}=1$.

 In this example, the velocity is vanishing at the equilibrium concentration $c_{\rm A}=1/6$.  At this special value where the equilibrium detailed balance conditions~(\ref{det-bal-12})-(\ref{det-bal-2}) are satisfied, the copolymer sequence forms a Bernoulli chain according to Eq.~(\ref{Bernoulli}) where the tip probabilities~(\ref{mu-eq}) take the values $\mu({\rm A})={\rm e}^{\varepsilon_{\rm A}}=1/3$ and $\mu({\rm B})={\rm e}^{\varepsilon_{\rm B}}=2/3$, since $\varepsilon_{\rm A}=\ln(1/3)$ and $\varepsilon_{\rm B}=\ln(2/3)$ for the parameter values~(\ref{Ex2}).  As a consequence of Eq.~(\ref{bar-mu-eq}), the bulk probabilities take the same values, which is confirmed by the value $\bar{\mu}({\rm A})=1/3$ observed in Fig.~\ref{fig2} at the concentration $c_{\rm A}=1/6$ where $v=0$.  Therefore, the equilibrium values of the free-energy driving force~(\ref{eps-eq}) and the sequence disorder~(\ref{D-eq}) are here equal to $D_{\rm eq}=-\varepsilon_{\rm eq}=\ln(3/2^{2/3})=0.63651$.  The affinity and the entropy production rate are thus vanishing at the concentration $c_{\rm A}=1/6$, as required if equilibrium is reached.
 
 In Fig.~\ref{fig3}, the free-energy driving force is vanishing ($\varepsilon=0$) at the concentration $c_{\rm A}=0.70485$, where $v=2.1147$, $A=D=0.68956$, and $d_{\rm i}S/dt=vD=1.4582$.  At this concentration, entropy is only produced due to sequence disorder. In the concentration range $1/6<c_{\rm A}<0.70485$, the free-energy driving force is negative, so that the copolymer growth is driven by the entropic effect of sequence disorder.  For $c_{\rm A}>0.70485$, the free-energy driving force is positive and the growth is driven by free energy.  For large values of the concentration, the free-energy driving force becomes dominant over sequence disorder: $\varepsilon\gg D$.  In this strongly irreversible regime where the attachment rates dominate, the entropy production rate can be evaluated with Eq.~(\ref{eps}) as
 \be
 \frac{d_{\rm i}S}{dt} \simeq v\, \varepsilon \simeq \sum_{m,i} \psi(i) \, k_{i,+m} \, c_m \ln \frac{k_{i,+m} \, c_m}{k_{i,-m}} \, ,
 \label{epr-large-conc}
 \ee
 in units where $k_{\rm B}=1$.  For $c_{\rm A}\gg c_{\rm B}$ in the example~(\ref{Ex2}), the entropy production rate increases as $d_{\rm i}S/dt\simeq (8c_{\rm A}/3)\ln(2c_{\rm A})$ with the concentration $c_{\rm A}$, as seen in Fig.~\ref{fig3}.

Figure~\ref{fig4} compares the normalized correlation function $C(n)=\Gamma(n)/\Gamma(0)$ of the quantity
\be
f(m) = \left\{
\begin{array}{cc}
1 & \mbox{if $m=$ A ,} \\
2 & \mbox{if $m=$ B ,}
\end{array}
\right.
\label{fn}
\ee
calculated in theory with Eq.~(\ref{correl}) in terms of the eigenvalue $\Lambda_2$ of the matrix~(\ref{R}) and in simulations with Eq.~(\ref{correl-app}), showing good agreement.  For $n=0$, the normalized correlation function is equal to unity.  For $n>0$, the correlation function drops to significantly lower values and it decays exponentially at the rate $\gamma=-\ln\Lambda_2$.  As observed in Fig.~\ref{fig4}, the correlation function decays faster and faster as the concentration $c_{\rm A}$ decreases, which is consistent with the absence of statistical correlations in the equilibrium limit where the sequences form Bernoulli chains.

\subsection{Example without thermodynamic equilibrium}

The parameters of this example are taken as
\bea
&& k_{1,+{\rm A}}=1 \, , \quad k_{1,+{\rm B}}=2 \, , \quad k_{2,+{\rm A}}=4 \, , \quad k_{2,+{\rm B}}=1 \, , \nonumber\\
&& k_{1,-{\rm A}}=5 \, , \quad k_{1,-{\rm B}}=3 \, , \quad k_{2,-{\rm A}}=6 \, , \quad k_{2,-{\rm B}}=2 \, , \nonumber\\
&&k_{12}=k_{21} = 1 \, , \qquad c_{\rm B}=1 \, .
\label{Ex4}
\eea
This other set of parameter values is not compatible with the existence of an equilibrium limit.

\begin{figure}[h]
\centerline{\scalebox{0.495}{\includegraphics{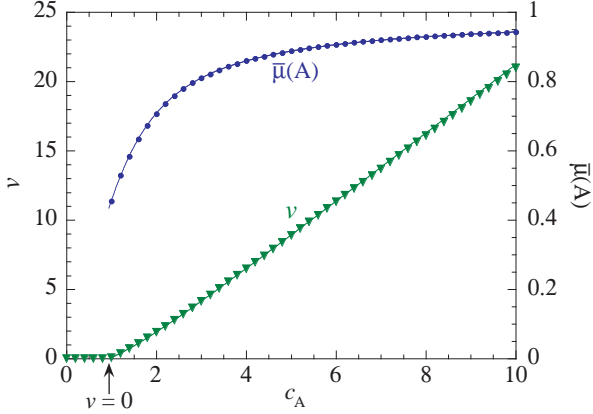}}}
\caption{Growth of a copolymer in the conditions (\ref{Ex4}): The mean growth velocity $v$ and the bulk probability $\bar{\mu}({\rm A})$ of the monomeric unit A, versus the concentrations $c_{\rm A}$.  The data points show the results of simulations with Gillespie's algorithm generating a sequence after $10^6$ jumps of the algorithm for every value of the concentration $c_{\rm A}$. The lines depict the theoretical predictions of Eqs.~(\ref{v-V-Psi}) and~(\ref{bulk-proba-1}).}
\label{fig5}
\end{figure}

\begin{figure}[h]
\centerline{\scalebox{0.495}{\includegraphics{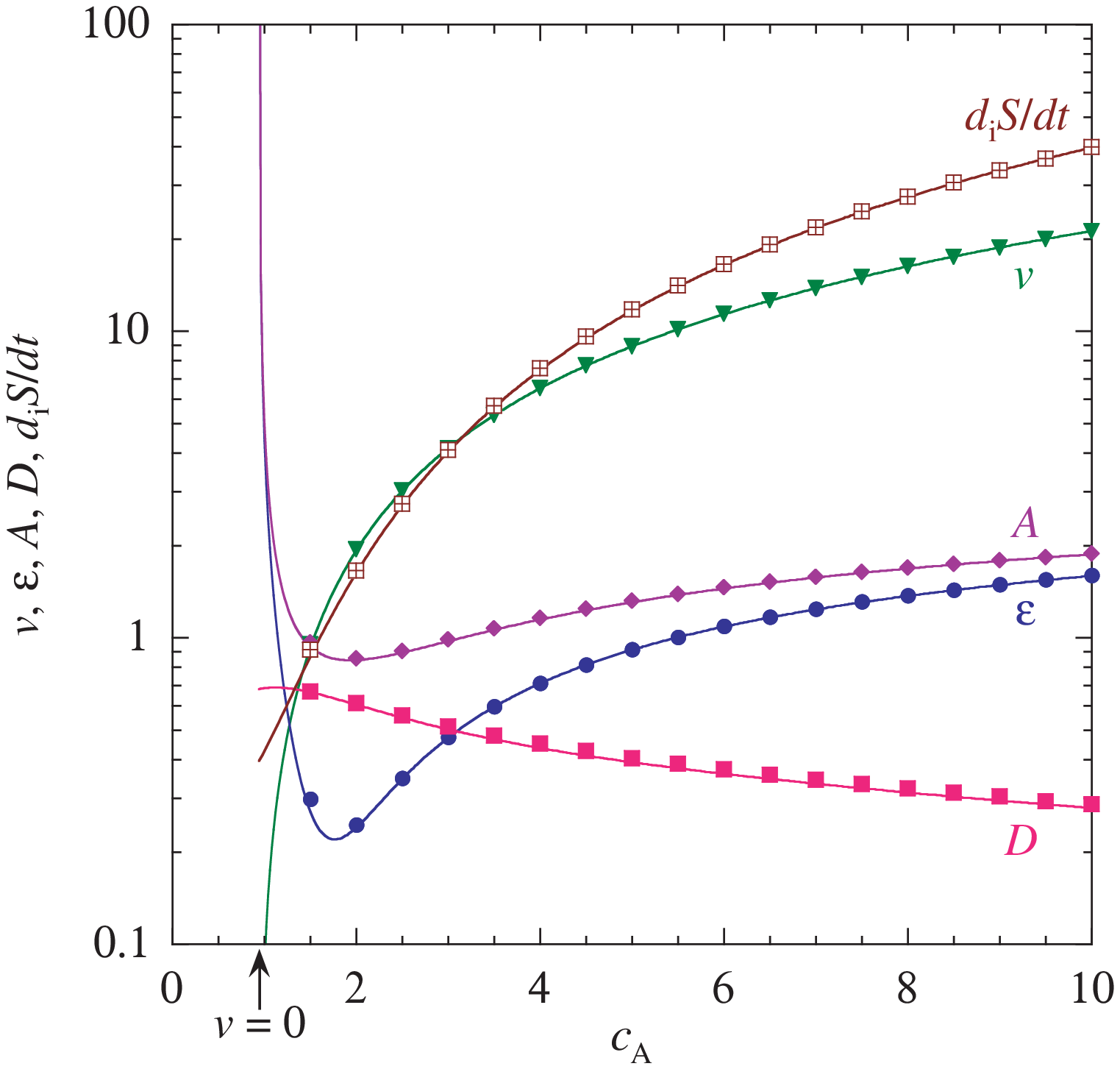}}}
\caption{Growth of a copolymer in the conditions (\ref{Ex4}): The mean growth velocity $v$, the free-energy driving force $\varepsilon$, the sequence disorder $D$, the affinity $A$, and the entropy production rate $d_{\rm i}S/dt$, versus the concentrations $c_{\rm A}$.  The data points are obtained with a statistics of $10^6$ sequences generated by Gillespie's algorithm running over the total time $t=200$.  
The lines show the theoretical predictions.}
\label{fig6}
\end{figure}

\begin{figure}[h]
\centerline{\scalebox{0.495}{\includegraphics{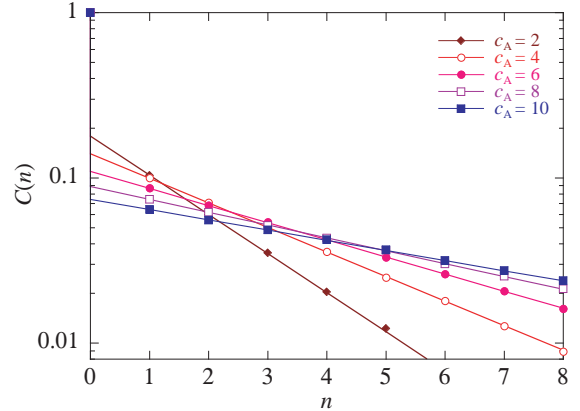}}}
\caption{Growth of a copolymer in the conditions (\ref{Ex4}):  The correlation function $C(n)$ versus $n$ for different values of the concentration $c_{\rm A}$.  The data points show the simulation results of Eq.~(\ref{correl-app}) using a long sequence generated by $10^8$ jumps of Gillespie's algorithm.  The total length of the sequence ranges from $L\simeq 1.6\times 10^7$ for the lowest concentration value $c_{\rm A}=2$ to $L\simeq 6.5\times 10^7$ for the largest one $c_{\rm A}=10$. The lines depict the theoretical predictions of Eq.~(\ref{correl}).}
\label{fig7}
\end{figure}

For this example, the mean growth velocity~(\ref{v-V-Psi}) and the bulk probability~(\ref{bulk-proba-1}) for the monomeric unit A are depicted in Fig.~\ref{fig5} as functions of the concentration $c_{\rm A}$ and the thermodynamic quantities in Fig.~\ref{fig6}.  Again, there is excellent agreement between the simulation results (data points) and the theoretical predictions (lines).  Here, the mean growth velocity vanishes at the critical concentration $c_{\rm A}=0.94233$.  However, the system remains out of equilibrium, as seen in Fig.~\ref{fig6} showing that, at the critical concentration,  the entropy production rate remains positive at the value $d_{\rm i}S/dt\simeq 0.39736$, while both the affinity $A$ and the free-energy driving force $\varepsilon$ diverge as $A\simeq \varepsilon \simeq 0.39736/v$ in consistency with the vanishing of the mean growth velocity.  At the critical concentration, the sequence disorder takes the value $D\simeq 0.68$.  Since the free-energy driving force remains positive in this example, the growth is always driven by free energy and there is here no regime of disorder-driven growth.

As in the previous example, the free-energy driving force becomes dominant over sequence disorder, $\varepsilon\gg D$, for large values of the concentration $c_{\rm A}$ where the attachment rates dominate.  In this strongly irreversible regime, the entropy production rate increases with the concentration $c_{\rm A}$ as $d_{\rm i}S/dt\simeq 2.5\, c_{\rm A}\ln(0.524\, c_{\rm A})$ according to Eq.~(\ref{epr-large-conc}), as observed in Fig.~\ref{fig6}.

For the example~(\ref{Ex4}), the normalized correlation function $C(n)=\Gamma(n)/\Gamma(0)$ of the quantity~(\ref{fn}) is shown in Fig.~\ref{fig7}, where the data points depict the values of Eq.~(\ref{correl-app}) for the simulations and the lines the theoretical functions calculated with Eq.~(\ref{correl}).  Here also, there is good agreement between theory and simulations with Gillespie's algorithm.  Again, the decay rate is slower at larger than smaller values of the concentration $c_{\rm A}$.

\section{Conclusion and perspectives}
\label{Conclusion}

In this paper, theory is developed for the kinetics and thermodynamics of multistate reversible copolymerization processes, in which the growing copolymer may undergo transitions between several reactive states controlling the monomeric attachment and detachment rates and these rates are supposed to be independent of previously incorporated monomeric units.  In this case, if the copolymer remained in a single reactive state, its sequences would form a Bernoulli chain.  However, as a consequence of the transitions between several reactive states, the growing copolymer sequences instead form a nonMarkovian chain. The probability distribution of the sequences are given in terms of products of matrices associated with every monomeric species and of size equal to the number of reactive states in the mechanism.  In general, the chain is nonMarkovian because the matrix products cannot be factorized as for Bernoulli or Markov chains.  The matrices of the theory also determine the mean growth velocity, the bulk probabilities of monomeric subsequences, the statistical correlation functions along the sequences, as well as the thermodynamic quantities.  In particular, the entropy production rate is shown to be given in terms of the mean growth velocity, the free-energy driving force, and the sequence disorder, confirming the results of Ref.~\onlinecite{AG08}.  

Two illustrative examples are used to compare theory with simulations using Gillespie's algorithm.  For the first example, the rate constants are compatible with the existence of a thermodynamic equilibrium limit.  In this example, the entropy production rate vanishes together with the mean growth velocity at the equilibrium value of the tuned monomeric concentration.  At this marginal concentration where the equilibrium detailed balance conditions are satisfied, the chain reduces to a Bernoulli chain.  Close to equilibrium, there exists a regime where the growth is driven by the entropy effect of sequence disorder, as predicted in Ref.~\onlinecite{AG08}.  For the second example, the rate constants are not compatible with the existence of an equilibrium limit.  Accordingly, the entropy production rate keeps a positive value at the critical concentration where the mean growth velocity vanishes and the free-energy driving force diverges together with the affinity.  In both examples, the correlation functions are observed to decay exponentially, as predicted by theory.  Excellent agreement is found between theory and simulations using Gillespie's algorithm.

All these results also concern multistate reversible homopolymerization processes yielding sequences with different tacticity.  Accordingly, they show that the theory of Coleman and Fox\cite{CF63JCP} can be extended from fully irreversible to reversible multistate mechanisms, allowing us to include the detachment of monomers beside their attachment in our models.  The present theory of multistate reversible copolymerization under low conversion conditions can be further extended to ultimate and penultimate multistate mechanisms where the attachment and detachment rates also depend on previously incorporated monomeric units.  For such other kinetics, the sequences form  Markov chains instead of Bernoulli chains,\cite{GA14,G16} if the copolymer remains in a single reactive state, but nonMarkovian chains if transitions occur between several reactive states.  The extension of the theory to template-directed copolymerization can also be considered.

\begin{acknowledgments}
This research is financially supported by the Universit\'e libre de Bruxelles (ULB) and the Fonds de la Recherche Scientifique~-FNRS under the Grant PDR~T.0094.16 for the project ``SYMSTATPHYS".
\end{acknowledgments}

\appendix

\section{The case of kinetics generating Bernoullian chains}
\label{AppA}

In this appendix, we show that we recover previous results for kinetics generating Bernoulli chains if either the transition rates $w_{1\to 2}$ or $w_{2\to 1}$ is vanishing.\cite{AG09}

If $w_{1\to 2}=0$, the chain stays in the reactive state $i=1$, so that $\psi(1)=1$ and $\psi(2)=0$.  Therefore, the mean growth velocity~(\ref{v-V-Psi}) is equal to $v=v_{11}+v_{21}$.  However, in Eq.~(\ref{eq-V}), the matrices ${\boldsymbol{\mathsf W}}_{-m}$ and ${\boldsymbol{\mathsf W}}_{+m}$ are diagonal and ${\boldsymbol{\mathsf W}}_0$ is upper triangular.  Consequently, the iteration~(\ref{eq-V}) converges towards an upper triangular matrix ${\boldsymbol{\mathsf V}}$, so that $v_{21}=0$.  The mean growth velocity is thus given by the solution $v=v_{11}$ of the self-consistent equation
\be
1 = \sum_m \frac{w_{i,+m}}{v+w_{i,-m}} ,
\label{self-consistent}
\ee
with $i=1$, as for the growth of a Bernoulli chain staying in this reactive state.

Similarly, if $w_{2\to 1}=0$, the chain stays in the reactive state $i=2$, so that $\psi(1)=0$ and $\psi(2)=1$.  In this case, the mean growth velocity~(\ref{v-V-Psi}) is equal to $v=v_{12}+v_{22}$.  In Eq.~(\ref{eq-V}), the matrix ${\boldsymbol{\mathsf W}}_0$ is now lower triangular, so that the iteration~(\ref{eq-V}) converges towards a lower triangular matrix ${\boldsymbol{\mathsf V}}$ and thus $v_{12}=0$.  Accordingly, the mean growth velocity is now given by the solution $v=v_{22}$ of the self-consistent equation~(\ref{self-consistent}) with $i=2$, as for the growth of a Bernoulli chain staying in this other reactive state.

Furthermore, if $w_{1\to 2}=w_{2\to 1}=0$, the matrix~(\ref{W0-dfn}) is equal to zero and the matrix equation~(\ref{eq-V}) reduces to
\be
{\boldsymbol{\mathsf V}}= {\boldsymbol{\mathsf V}}\cdot \sum_m ({\boldsymbol{\mathsf V}}+ {\boldsymbol{\mathsf W}}_{-m})^{-1} \cdot {\boldsymbol{\mathsf W}}_{+m} \, ,
\label{eq-V-B}
\ee
which is diagonal and decouples into the two previous self-consistent equations~(\ref{self-consistent}) for the mean growth velocities of the two types of copolymers, which are thus growing independently of each other.

\section{Numerical methods in theory}
\label{AppB}

If there are $I$ reactive states, the matrices are of size $I\times I$.  In the reported examples, we take $I=2$.
The central equation of the theoretical framework to be solved is Eq.~(\ref{eq-V}) for the velocity matrix.  
Its solution is obtained by iteration starting from an initial positive matrix ${\boldsymbol{\mathsf V}}_0$.
The right-hand side of Eq.~(\ref{eq-V}) gives the next iterate, which is reinserted in the right-hand side, and so on and so forth.
This iterative scheme converges towards the solution.  In the examples, $10^4$ iterations are used to obtain the solution.
Thereafter, the mean growth velocity can be calculated with Eq.~(\ref{v-V-Psi}) and the $M$ matrices ${\boldsymbol{\mathsf Y}}_m$ with Eq.~(\ref{eq-X-V}), giving the tip probabilities according to Eq.~(\ref{proba}).  In order to get the bulk probabilities and the correlation function, the eigenvalues and the eigenvectors of the matrix~(\ref{R}) are calculated.  The bulk probability $\bar{\mu}({\rm A})$ is thus obtained with Eq.~(\ref{bulk-proba-1}).

The free-energy driving force is given by Eq.~(\ref{eps}), the sequence disorder by Eq.~(\ref{D-dfn}), the affinity by Eq.~(\ref{Aff}), and the entropy production rate by Eq.~(\ref{epr-copolym}) in units where $k_{\rm B}=1$.  In order to obtain the sequence disorder~(\ref{D-dfn}), the Shannon sequence entropies
\be
D^{(l)} \equiv - \sum_{m_1\cdots m_l, i} \psi(m_1\cdots m_l, i) \, \ln \psi(m_1\cdots m_l, i)
\label{D-l-dfn}
\ee
are first calculated from the tip probabilities~(\ref{proba}).  The sequence disorder is thus given in principle by
\be
D =\lim_{l\to\infty} \left( D^{(l)}-D^{(l-1)} \right) ,
\label{D-diff}
\ee
and, in practice, with the value $D\simeq D^{(4)}-D^{(3)}$.

\section{Numerical methods in simulations}
\label{AppC}

The kinetic equations~(\ref{eq-i}) define a Markov jump stochastic process that can be exactly simulated using Gillespie's algorithm.\cite{G76,G77}  In the examples, we consider the growth of copolymer chains with $I=2$ reactive states and composed of $M=2$ monomeric units.  If the current state $\omega$ corresponds to the sequence $m_1\cdots m_l$ of length $l$ in the reactive state $i$, there are four possible random transitions to a new state $\omega'$ that may occur: (1) the attachment of the unit $m_{l+1}={\rm A}$ at the rate $w_{i,+{\rm A}}$; (2) the attachment of the unit $m_{l+1}={\rm B}$ at the rate $w_{i,+{\rm B}}$; (3) the detachment of the unit $m_l$ at the rate $w_{i,-m_l}$; and (4) the change of the reactive state from $i$ to $j$ at the rate $w_{i\to j}$.  The random waiting time $\Delta t(\omega\to\omega')$ before the jump is given by an exponential probability distribution of mean value $\tau=\left( w_{i,+{\rm A}} + w_{i,+{\rm B}} + w_{i,-m_l} + w_{i\to j}\right)^{-1}$.  The transition $\omega\to\omega'$ occurs with the probability $P(\omega\to\omega')=\tau W(\omega\to\omega')$, implying new values for the reactive state, the time, the length, and the cumulated free-energy driving force:
\bea
&& i \to j \, , \\
&& t \to  t + \Delta t (\omega\to \omega') \, , \\
&& l \to  l + \Delta l (\omega\to \omega') \, , \\
&& E \to E + \ln\left[W(\omega\to\omega')/W(\omega'\to\omega)\right] \, . \label{E-cumul}
\eea
After many jumps and the elapsed time reaching the value $t$, the mean growth velocity and the free-energy driving force are respectively evaluated by $v\simeq l/t$ and $\varepsilon\simeq E/t$.  The determination of the sequence disorder requires the computation of the tip probabilities $\psi(m_1\cdots m_l,i)$ with a large enough statistics, as specified in the figure captions.  In the simulations, the tip probabilities are obtained for $l=1,2,3$ and the value of sequence disorder is approximated by $D\simeq D^{(3)}-D^{(2)}$ in terms of the corresponding Shannon sequence entropies~(\ref{D-l-dfn}).  The affinity is thus calculated with Eq.~(\ref{Aff}) and the entropy production rate with Eq.~(\ref{epr-copolym}).  The correlation function is obtained for the function~(\ref{fn}) by using
\be
\Gamma(n) = \frac{1}{L} \sum_{k=1}^L \left[ f(m_k)-\langle f\rangle\right]\left[ f(m_{k+n})-\langle f\rangle\right] ,
\label{correl-app}
\ee
where $L$ is the total length of the long sequence used in the computation and $\langle f\rangle = (1/L)\sum_{k=1}^L  f(m_k)$ is the mean value of the function $f(m)$.


\end{document}